\let\orilabel\label
\documentclass[aps,pra,preprint,a4paper,showpacs,showkeys,superscriptaddress,nofootinbib]{revtex4-1}
\let\label\orilabel

\usepackage{latexsym}
\usepackage{amsmath,amssymb,mathrsfs}
\usepackage{mathtools}
\usepackage{graphicx}
\usepackage{subfigure}
\usepackage{xcolor}
\usepackage{physics}
\usepackage{cancel}
\usepackage{tikz}
\usepackage{multirow,tabularx}

\newcolumntype{C}[1]{>{\centering\arraybackslash}p{#1}}

\usepackage{acronym}

\usepackage{hyperref}     
\hypersetup{colorlinks,%
  citecolor=blue,%
  linkcolor=cyan,%
}

\usepackage[titletoc]{appendix}
\usepackage{enumerate}
\graphicspath{{./Figures/}}
\usetikzlibrary{decorations.pathmorphing,decorations.pathreplacing,decorations.shapes}
\acrodef{AdS}[AdS]{anti-de Sitter} 
\acrodef{BT}[BT]{Batalin-Tyutin}
\acrodef{JT}[JT]{Jackiw and Teitelboim}
\acrodef{WZ}[WZ]{Wess–Zumino}
\acrodef{BFT}[BFT]{Batalin–Fradkin–Tyutin}
\acrodef{ADT}[ADT]{Abbott-Deser-Tekin}
\acrodef{ADM}[ADM]{Arnowitt-Deser-Misner}

\begin{document}


\title{Extended symmetry of the Maxwell theory with a gauge coupling constant as a conserved charge}

\author{Sojeong Cheong}%
\email[]{jsquare@sogang.ac.kr}%
\affiliation{Center for Quantum Spacetime, Sogang University, Seoul 04107, Republic of Korea}%

\author{Myungseok Eune}%
\email[]{eunems@gmail.com}%
\affiliation{Center for Quantum Spacetime, Sogang University, Seoul
  04107, Republic of Korea}

\author{Wontae Kim}%
\email[]{wtkim@sogang.ac.kr}%
\affiliation{Center for Quantum Spacetime, Sogang University, Seoul 04107, Republic of Korea}%
\affiliation{Department of Physics, Sogang University, Seoul, 04107,
	Republic of Korea}%

\author{Mungon Nam}%
\email[]{clrchr0909@sogang.ac.kr}%
\affiliation{Center for Quantum Spacetime, Sogang University, Seoul
  04107, Republic of Korea}%
\date{\today}

\begin{abstract}
  It has been proposed that any coupling constant in a covariant action
  can be treated as a conserved charge by promoting the coupling constant to auxiliary fields, typically realized by a scalar field paired with a
  higher-form gauge field. However, the procedure may break local symmetries,
  which can be explicitly shown in a simpler setting such as Maxwell theory.
  The Hamiltonian analysis of Maxwell theory with the auxiliary fields
  reveals that some of the constraints are second-class. Applying the BFT formalism, we restore the broken local symmetries and obtain
  a fully symmetric action defined on an extended configuration space. Despite the restoration of the local symmetries, no additional conserved charges are associated with the recovered symmetries.
  Consequently, the original theory turns out to be the gauge-fixed version of the
  extended theory.
\end{abstract}
%



\maketitle


\raggedbottom

\section{Introduction}
\label{sec:introduction}

Since Bekenstein proposed that the entropy of a black hole is proportional to the area of
its event horizon~\cite{Bekenstein:1973ur,Bekenstein:1974ax}, the thermodynamics
of a wide class of black hole solutions has been extensively studied.
The thermodynamic quantities—such as mass, electric charge, and angular momentum—are
determined by physical charges identified with the integration constants appearing in the equations of motion rather than with the explicit parameters in the action.
Nevertheless, numerous works have suggested that parameters appearing in the gravitational action may themselves be interpreted as thermodynamic variables. For instance, the cosmological constant in AdS spacetime can be identified with a thermodynamic pressure, with its conjugate quantity interpreted as a thermodynamic volume~\cite{Henneaux:1984ji,Henneaux:1989zc,Teitelboim:1985dp}.
In this framework, the black hole mass is identified with the enthalpy rather than the internal
energy, and the resulting phase structure and thermodynamic behavior have been
extensively explored in
Refs.~\cite{Kastor:2009wy,Cvetic:2010jb,Dolan:2010ha,Dolan:2011xt,Frassino:2015oca,Kubiznak:2016qmn,Bhattacharya:2017hfj,Ortin:2021ade,Mann:2025xrb}. 
Fortunately, it is also possible to promote the cosmological constant to an integration
constant appearing in the equations of motion by introducing an auxiliary field~\cite{Aurilia:1980xj,Duff:1980qv,Hawking:1984hk,Bousso:2000xa,Chernyavsky:2017xwm,Hajian:2021hje}.
More recently, it has been demonstrated that essentially any coupling constant
in an action can be elevated to a conserved charge arising from a
local gauge symmetry by introducing a scalar field together
with a $(D-1)$-form gauge
field~\cite{Meessen:2022hcg,Zatti:2023oiq,Ballesteros:2023muf,Hajian:2023bhq}.
Along the lines of Refs.~\cite{Meessen:2022hcg,Zatti:2023oiq,Ballesteros:2023muf,Hajian:2023bhq}, the gauge coupling in the Einstein-Maxwell theory can be promoted to a dynamical field.
If one takes the flat limit, the dynamical gauge coupling still remains, and the resulting theory would be described by Maxwell theory with the dynamical gauge coupling.

To raise the issue in the simplest setting, we will consider Maxwell theory with the gauge coupling promoted to a dynamical field. Then, we perform a Hamiltonian analysis in order to investigate the symmetry structure of the theory.
The analysis shows that a subset of the constraints becomes second-class~\cite{dirac2001lectures}, which means that certain local symmetries
are broken upon introducing auxiliary fields.
If such an analysis reveals that certain symmetries are indeed broken, they can be restored by appropriate methods. In fact, there are two primary approaches for converting second-class constraint systems into first-class ones.
The first approach is to add the Wess-Zumino action~\cite{Wess:1971yu} to cancel the anomaly
responsible for the second-class constraint algebra in the path-integral formalism~\cite{Faddeev:1986pc,Babelon:1986sv,Harada:1986wb,Miyake:1987dk}.
The second is to employ the Batalin–Fradkin-Tyutin Hamiltonian method in the Hamiltonian framework~\cite{Batalin:1986aq,Batalin:1986fm,Batalin:1991jm},
which has been applied to Chern-Simons field theories~\cite{Banerjee:1993pm,Kim:1994np}, anomalous gauge theories~\cite{Fujiwara:1989ia,Fujiwara:1990rx,Kim:1992ey,Banerjee:1993pj},
and chiral boson theories~\cite{McClain:1990sx,Wotzasek:1990zr,Amorim:1994np,Amorim:1994ft,Kim:2006za}.

In this paper, we consider the Maxwell action in which the electromagnetic field is coupled
to a scalar–vector pair~\cite{Meessen:2022hcg,Zatti:2023oiq,Ballesteros:2023muf,Hajian:2023bhq}.
We explicitly derive the quasi-local conserved charges $q$ and $e$
related to the two types of gauge symmetries parameterized by $\lambda$ and $\Lambda^{\mu\nu}$.
Using a detailed Dirac analysis of the constraint structure~\cite{dirac2001lectures},
we identify a set of second-class
constraints. We then restore the broken local symmetries by converting
the second-class constraints into first-class ones within the \ac{BFT} embedding
formalism, thereby obtaining a fully symmetric action defined on an extended configuration space.
Finally, we find the explicit form of the symmetry transformations for all first-class constraints,
where the original Maxwell theory is interpreted as the gauge-fixed version of the extended theory by choosing the unitary gauge.

The paper is organized as follows. In Sec.~\ref{sec:e:conserved.charge}, we
present the Maxwell action with a scalar–vector pair and obtain the Noether charges associated with the global parts of the local gauge symmetries.
In Sec.~\ref{sec:Dirac}, we classify six first-class and two second-class constraints using the Dirac method.
In Sec.~\ref{sec:BFT}, the broken local symmetries due to the two second-class constraints
are restored using the \ac{BFT} embedding method.
In Sec.~\ref{sec:new.symmetry}, we derive local symmetry transformations for the new first-class constraints as well as the original first-class constraints.
Conclusion and discussion are given in Sec.~\ref{sec:conclusion}.

\section{Coupling constant as a conserved charge}
\label{sec:e:conserved.charge}
The coupling constant in Maxwell’s electromagnetic theory can be
promoted to a dynamical field through the action~\cite{Meessen:2022hcg,Zatti:2023oiq,Ballesteros:2023muf,Hajian:2023bhq}
\begin{equation}
  \label{eq:action}
  S[A_\mu, \sigma, \chi_\mu] =\int \dd[4]{x}\left[ -\frac{1}{4}\sigma\left( F_{\mu\nu}F^{\mu\nu}
      - \partial_{\mu}\chi^\mu \right) \right],
\end{equation}
where $F_{\mu\nu} = \partial_{\mu} A_{\nu} - \partial_{\nu} A_{\mu}$ denotes the electromagnetic field strength tensor, $\sigma$ is a scalar field, and
$\chi^{\mu}$ is an auxiliary vector field with the Minkowski metric $\eta_{\mu\nu} = \operatorname{diag}(+1,-1,-1,-1)$.
In the action~\eqref{eq:action}, the gauge coupling is effectively replaced
by the scalar $\sigma$ along with the auxiliary field $\chi^{\mu}$.
One readily verifies that the action \eqref{eq:action} is invariant
under two independent local gauge transformations:
\begin{equation}
  \label{eq:gauge transform}
  \delta A_{\mu} = \partial_{\mu} \lambda, \quad
  \delta\chi^\mu = \partial_{\nu}\Lambda^{\mu\nu},
  \end{equation}
where $\lambda$ is a local gauge parameter and $\Lambda^{\mu\nu}$ is an
arbitrary antisymmetric tensor gauge parameter.

Varying the
action~\eqref{eq:action} with respect to the fields yields
\begin{equation}
  \label{eq:var action}
  \delta S = \int \dd[4]{x}\left[ -\mathcal{E}^\mu\delta A_{\mu} -
    \mathcal{E}^{\chi}_\mu \delta \chi^\mu - \mathcal{E}^\sigma \delta \sigma +
    \partial_{\mu}\Theta^{\mu}\left( \delta A, \delta \chi \right)\right],
\end{equation}
where $ \mathcal{E}^\mu = \partial_{\nu}\left( \sigma F^{\mu\nu} \right),\quad
  \mathcal{E}_{\mu}^{\chi}= \frac{1}{4}\partial_{\mu}\sigma,\quad
  \mathcal{E}^\sigma = \frac{1}{4}\left(
    F_{\mu\nu}F^{\mu\nu}-\partial_{\mu}\chi^\mu \right)$,
and the boundary term is
$  \Theta^\mu\left( \delta A, \delta \chi \right) = -\sigma F^{\mu\nu}\delta
  A_{\nu} + \frac{1}{4}\sigma \delta \chi^\mu$.
From the equations of motion, $\mathcal{E}^\mu = 0$, $\mathcal{E}^\chi_\mu = 0$,
and $\mathcal{E}^\sigma = 0$,  a static
solution in spherical coordinates can be found as
\begin{equation}
  \label{solution:fields}
  A_{t} = \frac{q}{4\pi r},\quad \sigma = \frac{1}{e^2},\quad \chi^r =
  \frac{q^2}{8\pi^2 r^3},
\end{equation}
where $q$ and $e^2$ are integration
constants.
Substituting the gauge transformations \eqref{eq:gauge transform} into Eq.~\eqref{eq:var action}, one obtains
\begin{equation}
  \label{eq:Noether current}
  0 = \int \dd[4]{x}\partial_{\mu}\left[ -\mathcal{E}^{\mu}\lambda +
    \mathcal{E}^{\chi}_\nu \Lambda^{\mu\nu} + \Theta^{\mu}\left( \lambda,
      \Lambda \right)\right],
\end{equation}
where $\Theta^\mu (\lambda, \Lambda)$ denotes the boundary term related to the symmetry variations. From Eq.~\eqref{eq:Noether current}, we define the off-shell
Noether current
as
$  J^\mu(\lambda,\Lambda) = -\mathcal{E}^{\mu}\lambda + \mathcal{E}^{\chi}_\nu
  \Lambda^{\mu\nu} + \Theta^{\mu}\left( \lambda, \Lambda \right)$,
where $\partial_{\mu}J^\mu = 0$.
Thus, in virtue of the Poincar\'e
lemma, there exists an off-shell potential $K^{\mu\nu}$ satisfying
$J^{\mu} = \partial_{\nu}K^{\mu\nu}$, which leads to
\begin{equation}
  \label{eq:Noether potential K}
  K^{\mu\nu}\left( \lambda, \Lambda \right) = -\sigma \lambda F^{\mu\nu} +
  \frac{1}{4}\sigma \Lambda^{\mu\nu}.
\end{equation}
For the global part of the symmetries, we choose the parameters
$\lambda = e^2$ and $\Lambda^{tr} = \frac{e^3}{\pi r^2}$.
Using Eqs.~\eqref{solution:fields} and \eqref{eq:Noether potential K}, one then obtains the Noether charges:
\begin{align}
  Q\left[ \lambda,0 \right]
  &= \oint \dd[2]{x_{\mu\nu}} K^{\mu\nu}(\lambda,0) = - \oint
    \dd{\theta} \dd{\phi} r^{2}\sin\theta F^{tr} = q,\\
  Q\left[ 0,\Lambda \right]
  &= \oint \dd[2]{x_{\mu\nu}} K^{\mu\nu}(0,\Lambda) = \frac{e}{4} \oint
    \dd{\theta} \dd{\phi} r^{2}\sin\theta\Lambda^{tr} = e.
\end{align}
Thus, the conserved electric charge $q$ and the Noether charge $e$ are arising from the gauge symmetries implemented by $\lambda$ and  $\Lambda^{\mu\nu}$, respectively.

\section{Constraint structures}
\label{sec:Dirac}

Hereafter, for notational clarity, we relabel the original fields in the
action~\eqref{eq:action} as
$S[A_\mu, \sigma, \chi_\mu] \to S^{(0)}[A_\mu^{(0)}, \sigma^{(0)}, \chi_{\mu}^{(0)}]$.
In order to study the symmetry structure of the system, we employ the Dirac
method~\cite{dirac2001lectures}. From
the action~\eqref{eq:action}, the canonical momenta for $\sigma^{(0)}$,
$\chi_\mu^{(0)}$, and $A_\mu^{(0)}$ are obtained as
$\pi_\sigma^{(0)} =0,~\pi_\chi^{(0)\mu} = \frac14 \eta^{\mu 0}\sigma^{(0)},~\pi_A^{(0)\mu}
= \sigma^{(0)} F^{\mu 0}$.  Accordingly, the primary constraints can be identified as
\begin{gather}
  \Omega_1 = \pi_\sigma^{(0)}  \approx 0,\quad \Omega_2 = \pi_{\chi}^{(0)0} -
  \frac14 \sigma^{(0)}  \approx 0,\quad
  \Omega_3^i = \pi_{\chi}^{(0)i} \approx
  0,\quad \Omega_4 = \pi_{A}^{(0)0} \approx 0.\label{constraints:primary:1st.class}
\end{gather}
Performing the Legendre transformation of the system~\eqref{eq:action} in the presence of the primary constraints~\eqref{constraints:primary:1st.class},
one obtains the primary Hamiltonian as
\begin{align}
  H_{\mathrm{p}} = \int \dd[3]{\mathbf{x}} \bigg[ &\frac{1}{2\sigma^{(0)}}
        \pi_{A}^{(0)i}\pi_{A}^{(0)i} + \frac14 \sigma^{(0)} (F^{(0)}_{ij}F^{(0)ij} -
                                                    \partial_{i}\chi^{(0)i} )
  - A_0^{(0)}\partial_{i}\pi_{A}^{(0)i}  \notag \\
  &+ u_1 \Omega_1 + u_2 \Omega_2 +
  u_{3i} \Omega_3^i + u_4 \Omega_4 \bigg], \label{H.p}
\end{align}
where $u$'s are the Lagrange multipliers.  Note that the constraints
$\Omega_1$ and $\Omega_2$ in
Eq.~\eqref{constraints:primary:1st.class} are second-class, while
$\Omega_3^i$ and $\Omega_4$ are first-class.  The time evolution of
${\Omega}_1$ and
${\Omega}_2$ fixes the Lagrange multipliers as
$u_2 = -\frac{2}{(\sigma^{(0)})^{2}} \pi_{A}^{(0)i} \pi_{A}^{(0)i} + F^{(0)}_{ij} F^{(0)ij} -
\partial_i \chi^{(0)i}$ and $u_1 = 0$, respectively.  Additionally, the time
evolution of $\Omega_3^i$ becomes
$\dot{\Omega}_3^{i} = \frac{1}{4} \partial_i \sigma^{(0)} =
\hat{\Omega}_{5i} \approx 0$. For convenience, we redefine
$\Omega_{5i} = \hat{\Omega}_{5i} + \partial_i \Omega_2 =
\partial_i \pi_\chi^{0} \approx 0$. The time evolution of $\Omega_{5i}$ does not give rise to any further constraints.  Next, the time evolution
of $\Omega_4$ generates the Gauss-law constraint of
$\Omega_6 = \partial_i \pi_{A}^{(0)i} \approx 0$.  As a result, $\Omega_3^i$,
$\Omega_4$, $\Omega_{5i}$, and $\Omega_6$ are first-class, and
$\Omega_1$ and $\Omega_2$ are second-class. Explicitly, the second-class
constraint algebra can be written as
\begin{equation}
  \{\Omega_a (x),\Omega_b (y)\} = \frac{1}{4} \epsilon_{ab}
  \delta^3 (\mathbf{x}-\mathbf{y}),\quad a=1,2, \label{secondclass}
\end{equation}
where $\epsilon_{ab}$ is the Levi-Civita symbol with $\epsilon_{12} = +1$.
Consequently, despite the presence of the local gauge symmetries~\eqref{eq:gauge transform}, the two primary constraints constitute a set of second-class constraints,
which means that additional local symmetries are broken in the system.

\section{Batalin-Fradkin-Tyutin embedding}
\label{sec:BFT}

In this section, we restore the broken local symmetries by
converting the two second-class constraints $\Omega_a$ into first-class
constraints, thereby obtaining a fully gauge-invariant action within the
\ac{BFT} formalism~\cite{Batalin:1986aq,Batalin:1986fm,Batalin:1991jm}.
We begin by extending the original phase space
by introducing two auxiliary fields $\varphi_a^{(1)}$, one for each
second-class constraint.
The auxiliary fields are assumed to satisfy the Poisson bracket
\begin{align}
  \{ \varphi_a^{(1)} (x), \varphi_b^{(1)} (y) \} = - \epsilon_{ab} \delta^3(\mathbf{x}-\mathbf{y}). \label{PB:sigma.a.1}
\end{align}
In the extended phase space, the constraints $\tilde\Omega_a$ are assumed to be
\begin{align}
  \tilde\Omega_a = \Omega_a + \sum_{k=1}^{\infty}
  \omega_a^{(1,k)}, \label{Omega.a.0:new}
\end{align}
where $\Omega_a$ are the original second-class constraints satisfying Eq.~\eqref{secondclass}.
On general grounds, each correction term $\omega_a^{(1,k)}$ is supposed to be of order $k$ in the auxiliary fields, $i.e.$, proportional to $(\varphi_a^{(1)})^k$.
However, it is sufficient to retain only the first-order correction as
$\omega_a^{(1,1)} = \int \dd[3]{\mathbf{y}} X_{ab}(x,y)\, \varphi_b^{(1)}(y) $
so that $\omega_a^{(1,k)} = 0$ for $k\ge 2$.  The requirement of the involution condition of $\{ \tilde{\Omega}_a (x), \tilde{\Omega}_b (y) \} \approx 0 $ leads to
the simplest solution for $X_{ab}$, given by
\begin{align}
    X_{ab}(x,y)
  = \begin{pmatrix}
    -1 & 0 \\
    0 & - \frac14
  \end{pmatrix}
        \delta^3(\mathbf{x}-\mathbf{y}). \label{X:ab}
\end{align}
Redefining the auxiliary fields as $\varphi_2^{(1)} = \sigma^{(1)}$ and
$\varphi_1^{(1)}=\pi_\sigma^{(1)}$ in Eq.~\eqref{PB:sigma.a.1}, one can regard $\pi_\sigma^{(1)}$ as the
conjugate momentum of the field $\sigma^{(1)}$, satisfying
$ \{ \sigma^{(1)}(x), \pi_\sigma^{(1)} (y) \} = \delta^3(\mathbf{x}-\mathbf{y})$.
With these definitions, the constraints~\eqref{Omega.a.0:new} can be written as
\begin{align}
  \tilde\Omega_1 &= \pi_\sigma^{(0)} - \pi_\sigma^{(1)}\approx
                         0, \quad
  \tilde\Omega_2 = \pi_\chi^{(0)0} - \frac14 (\sigma^{(0)} +
                         \sigma^{(1)}) \approx 0. \label{Omega.a.0:1.modified}
\end{align}
For later convenience, we denote the collection of all first-class constraints by $\Omega_\alpha= \{\tilde{\Omega}_{a},~\Omega_{3}^{i},~\Omega_4,~\Omega_{5}^{i}, ~\Omega_6\}$.

Next, the extended canonical
Hamiltonian~\cite{Batalin:1986aq,Batalin:1986fm,Batalin:1991jm}
\begin{align}
  \tilde{H}_{\mathrm{c}}^{(0)} &= H_{\mathrm{c}}^{(0)} +  \sum_{k=1}^\infty
                                 h^{(1,k)}\label{H.c:1}
\end{align}
can be determined such that it satisfies
$\{ \tilde{\Omega}_a, \tilde{H}_{\mathrm{c}}^{(0)} \} \approx 0$, where
$H_{\mathrm{c}}^{(0)} $ denotes the canonical Hamiltonian for the
action~\eqref{eq:action} and $h^{(1,k)} \sim (\varphi_a^{(1)})^k$.
Following the
\ac{BFT} formalism, $h^{(1,n)}$ in Eq.~\eqref{H.c:1} is constructed as
\begin{align}
  h^{(1,n)} &= - \frac{c_n}{n!} \int \dd[3]{\mathbf{x}} \dd[3]{\mathbf{y}}
            \dd[3]{\mathbf{z}} \varphi_a^{(1)} (x) \gamma_{ab} (x,y)
            X_{bc} (y,z) \mathcal{G}_c^{(n-1)} (z), \label{def:h.1.n}
\end{align}
where $c_n$ is a constant and
$\mathcal{G}_a^{(n)}$ is defined recursively by
\begin{align}
  \mathcal{G}_a^{(0)} &= \{ \Omega_a,  H_{\mathrm{c}}^{(0)} \},  \label{G.a:0}\\
  \mathcal{G}_a^{(1)} &= \{ \Omega_a, h^{(1,1)} \} + \{
                        \omega_a^{(1,1)}, H_{\mathrm{c}}^{(0)} \}_{(1)}, \label{G.a:1}\\
  \mathcal{G}_a^{(n)} &= \sum_{m=0}^n \{ \omega_a^{(1,n-m)}\!,\! h^{(1,m)} \}_{(0)}
                        \!+\! \sum_{m=0}^{n-2} \{ \omega_a^{(1,n-m)}\!,\! h^{(1,m+2)}
                        \}_{(1)} \!+\! \{ \omega_a^{(1,n+1)}\!,\! h^{(1,1)}
                        \}_{(1)} ~~~ (n\ge 2) \label{G.a:n}
\end{align}
with the conventions $\omega_a^{(1,0)} = \Omega_a$ and
$h^{(1,0)} = H_{\mathrm{c}}^{(0)}$.  The subscripts $(0)$ and $(1)$ on
the Poisson brackets in Eqs.~\eqref{G.a:1} and \eqref{G.a:n} indicate that the
brackets are evaluated only over the original fields and the auxiliary fields,
respectively.
At the zeroth order in
$\varphi_a^{(1)}$, $h^{(1,1)}$ is obtained from Eq.~\eqref{def:h.1.n} as follows:
\begin{align}
  h^{(1,1)} &= \int \dd[3]{\mathbf{x}} \left[
              \frac{ (-\sigma^{(1)})}{2 (\sigma^{(0)})^{2}}
              \pi_{A}^{(0)i} \pi_{A}^{(0)i} + \frac14 \sigma^{(1)} \left(F^{(0)}_{ij}
              F^{(0)ij} - \partial_i \chi^{(0)i}\right) \right]. \label{h.1:result}
\end{align}
The Poisson bracket $ \{ \tilde{\Omega}_a, H_c^{(0)} \} $ partially cancels $ \{ \tilde{\Omega}_a$, $h^{(1,1)} \} $, leaving
an undesired term proportional to $ \varphi_2^{(1)} = \sigma^{(1)}$.
To eliminate the term, we should consider $h^{(1,2)}$; however, this process produces an additional term as well.
Thus, this procedure should be iterated infinitely.
As a result, one finds that $h^{(1,k)}$ takes the form of
\begin{align}
  h^{(1,k)} &= \int \dd[3]{\mathbf{x}} \left[
            \frac{ (-\sigma^{(1)})^k}{2 (\sigma^{(0)})^{k+1}}
            \pi_{A}^{(0)i} \pi_{A}^{(0)i} + \frac14 \delta_{1k} \sigma^{(1)} \left(F^{(0)}_{ij}
            F^{(0)ij} - \partial_i \chi^{(0)i}\right) \right], \label{h.n:result}
\end{align}
and thus the extended canonical Hamiltonian~\eqref{H.c:1} is obtained as
\begin{align}
  \tilde{H}_{\mathrm{c}}^{(0)} \!=\!
  & \int \!\dd[3]\!{\mathbf{x}} \left[ \frac{1}{2(\sigma^{(0)} \!+\! \sigma^{(1)} )} \pi_{A}^{(0)i}
    \pi_{A}^{(0)i} \!+\! \frac14 (\sigma^{(0)}  \!+\! \sigma^{(1)} )  \left(F^{(0)}_{ij}
    F^{(0)ij} \!-\! \partial_i \chi^{(0)i} \right) \!-\! A_0^{(0)}
    \partial_i \pi_{A}^{(0)i} \right]\!.\! \label{H.c:1:final}
\end{align}
Indeed, one can directly verify the relation $\{ \tilde{\Omega}_a, \tilde{H}_{\mathrm{c}}^{(0)} \} \approx 0$ by using Eqs.~\eqref{Omega.a.0:1.modified} and~\eqref{H.c:1:final}, as required.

We are now in a position to derive the
action from the extended canonical Hamiltonian~\eqref{H.c:1:final}, which must be compatible with all constraints $\Omega_\alpha$.
The partition function in the extended phase space is written as
\begin{align}
  Z &= \int \mathcal{D} A_\mu^{(0)} \mathcal{D}
      \sigma^{(0)}\mathcal{D}
      \sigma^{(1)} \mathcal{D} \chi_\mu^{(0)} \mathcal{D} \pi_A^{(0)\mu}
      \mathcal{D} \pi_\sigma^{(0)} \mathcal{D} \pi_\sigma^{(1)} \mathcal{D} \pi_\chi^{(0)\mu}
      \delta[\Omega_{\alpha}]\delta[\Gamma_{\alpha}]\det|\{
      \Omega_{\beta}, \Gamma_{\gamma} \}| \, e^{i\tilde{S}_c^{(0)}}, \label{def:Z}
\end{align}
where $  \tilde{S}^{(0)}_c = \int \dd[4]{x} \left[\pi_A^{(0)\mu} \dot{A}_\mu + \pi_\chi^{(0)\mu}
  \dot{\chi}_\mu^{(0)} +  \pi_\sigma^{(0)} \dot{\sigma}^{(0)}
 + \pi_\sigma^{(1)} \dot{\sigma}^{(1)}  \right] - \int \dd{t} \tilde{H}_{\mathrm{c}}^{(0)} $
and $\Gamma_{\alpha}$ denotes the gauge fixing condition.
In the partition function~\eqref{def:Z}, we integrate out the momenta by taking into account the delta functionals of
the constraints in the measure. Integrating out the variables
$\pi_\sigma^{(0)}$,~$\pi_\sigma^{(1)}$,~and $\pi_\chi^{(0)0}$,
we obtain a manifestly covariant term $\delta\!\left[ \partial_\mu (\sigma^{(0)} + \sigma^{(1)}) \right]$,
which can be exponentiated in the action by introducing a new auxiliary field $\chi^{(1)\mu}$.
In addition, after integrating out $\pi_A^{(0)\mu}$ and $\pi_\chi^{(0)i}$,
the partition function can be expressed as
\begin{align}
  Z =  \int \mathcal{D} A_\mu^{(0)} \mathcal{D}
  \sigma^{(0)}  \mathcal{D} \sigma^{(1)} \mathcal{D}  \chi_\mu^{(0)} \mathcal{D}
  \chi_\mu^{(1)}  \delta[\Gamma_{\alpha}]\det|\{ \Omega_{\beta},
  \Gamma_{\gamma} \}|\, e^{i \tilde{S}^{(0)}}, \label{Z:S.eff:1}
\end{align}
where the extended action takes the form:
\begin{align}
  \tilde{S}^{(0)}
  = \int \dd[4]{x} \bigg[&- \frac14 \sigma^{(0)} \left( F^{(0)}_{\mu\nu}
     F^{(0)\mu\nu} - \partial_\mu \chi^{(0) \mu} \right) - \frac14 \sigma^{(1)} \left( F^{(0)}_{\mu\nu}
     F^{(0)\mu\nu} - \partial_\mu \chi^{(1) \mu} \right)  \notag \\
  &+ \frac14 \sigma^{(1)}
     \partial_\mu \chi^{(0) \mu} - \frac14 \chi^{(1)
     \mu} \partial_\mu \sigma^{(0)}\bigg].
      \label{S.1:modified}
\end{align}
It is worth noting that the auxiliary field $\chi^{(1)\mu}$ is explicitly introduced at the level of the action, and may therefore deform the first-class constraint structure~\eqref{Omega.a.0:1.modified}. To make this point explicit, we derive the following constraints from the action~\eqref{S.1:modified}:
\begin{align}
  \Psi_1 &= \pi_\sigma^{(1)}  \approx 0,
                 & &\Psi_2 = \pi_\chi^{(1)0}  - \frac14 \sigma^{(1)}  \approx
                 0, \label{Omega.2:1}\\
  \Psi_3 &= \pi_\sigma^{(0)} - \pi_\sigma^{(1)} +  \frac14
           \chi^{(1)0} \approx 0,
           & &\Psi_4 = \pi_\chi^{(0)0} - \frac14
                 \sigma^{(0)} - \pi_\chi^{(1)0} \approx 0,
                 \label{Omega.1:1}\\
  \Psi_{5}^{i} &=   \pi_\chi^{(0)i} \approx 0, \qquad\quad
                 \Psi_{6}^{i} = \pi_\chi^{(1)i} \approx 0,
                 &   &\Psi_7 = \pi_A^{(0)0} \approx 0, \label{Omega.p.c.others:1} \\
  \Psi_{8i} &= \partial_i \pi_\chi^{(0)0} \approx 0,
              & &\Psi_9 = \partial_i \pi_A^{(0)i} \approx 0. \label{constraints:secondary:1st}
\end{align}
The constraints in Eqs.~\eqref{Omega.2:1}-\eqref{Omega.p.c.others:1}
constitute primary constraints, while those in Eq.~\eqref{constraints:secondary:1st} are secondary constraints.
Moreover, the constraints in Eqs.~\eqref{Omega.1:1}-\eqref{constraints:secondary:1st}
form a first-class set, whereas the two constraints in Eq.~\eqref{Omega.2:1} are second-class.
Although the covariant
action~\eqref{S.1:modified} can be obtained through the new auxiliary field
$\chi^{(1)\mu}$, it nevertheless gives rise to the second-class constraint algebra:
\begin{equation}
    \{\Psi_a(x), \Psi_b(y)\} = \frac{1}{4} \epsilon_{ab}
    \delta^3(\mathbf{x} - \mathbf{y}), \label{second.class:1}
\end{equation}
which is identical to the algebraic relation~\eqref{secondclass}.  Thus, the \ac{BFT} embedding must be iterated once more and, in fact, continued indefinitely, following the same procedure as before.
After some tedious calculations, we obtain the modified action as
\begin{align}
    \tilde{S} \!=\!
  &\int \!\dd[4]{x}\! \left[- \frac14 \sum_{n=0}^\infty \left( \sigma^{(n)} \left( F^{(0)}_{\mu\nu}
    F^{(0)\mu\nu} \!-\! \partial_\mu \chi^{(n)\mu} \right)
    \!+\! \sum_{m=n+1}^\infty \!\left( \chi^{(m)\mu} \partial_\mu
    \sigma^{(n)} - \sigma^{(m)} \partial_\mu \chi^{(n)\mu} \right)
    \!\right) \!\right]\!,\! \label{S:final:1st.class}
\end{align}
where an infinite set of auxiliary fields is introduced to render all constraints first-class. The feature is closely analogous to that encountered in chiral boson theories~\cite{McClain:1990sx,Wotzasek:1990zr,Amorim:1994np,Amorim:1994ft,Kim:2006za} which need to be iterated infinitely.

Lastly, we examine the validity of Eq.~\eqref{S:final:1st.class} by verifying that the resulting action describes a
first-class constraint system. The canonical momenta conjugate to the fields
$A_\mu^{(0)}$, $\sigma^{(n)}$, and $\chi_\mu^{(n)}$ are obtained as
\begin{align}
  \pi_A^{(0)\mu} &=  \sum_{m=0}^\infty \sigma^{(m)}\, F^{(0)\mu 0}, \quad
  \pi_\sigma^{(n)} = - \frac14 \sum_{m=n+1}^\infty \chi^{(m)0}, \quad
  \pi_\chi^{(n)\mu} = \frac14 \eta^{\mu 0}\,  \sum_{m=n}^\infty
              \sigma^{(m)}, \label{momemta:infinite.aux.fileds}
\end{align}
where $n\ge 0$.
The momenta~\eqref{momemta:infinite.aux.fileds} lead to the following set of primary
constraints:
\begin{align}
  \Phi_1^{(n)}
  &= \pi_\sigma^{(n)} - \pi_\sigma^{(n+1)} + \frac14
    \chi^{(n+1)0}\approx 0, \label{Omega.1:n} \\
  \Phi_2^{(n)}
  &=  \pi_\chi^{(n)0} - \frac14 \sigma^{(n)}- \pi_\chi^{(n+1)0}
    \approx 0, \label{Omega.2:n}\\
  \Phi_{3}^{(n)i} &=   \pi_\chi^{(n)i} \approx 0, \label{Omega.3i:n}\\
  \Phi_4 &= \pi_A^{(0)0} \approx 0. \label{Omega.4:n}
\end{align}
Performing the Legendre transformation of the
action~\eqref{S:final:1st.class}, we obtain the canonical Hamiltonian as
\begin{align}
  \tilde{H}_{\mathrm{c}}
  &= \int \dd[3]{\mathbf{x}} \bigg[
  \frac{1}{2\tilde\sigma} \pi_{A}^{(0)i}
    \pi_{A}^{(0)i} + \frac14 \sum_{n=0}^\infty \bigg( \sigma^{(n)} (F^{(0)}_{ij}
    F^{(0)ij} -  \partial_i \chi^{(n)i} )  \notag \\
  &\qquad\qquad + \sum_{m=n+1}^\infty
    \left(\chi^{(m)i} \partial_i \sigma^{(n)} - \sigma^{(m)} \partial_i
    \chi^{(n)i} \right) \bigg)  - A_0^{(0)} \partial_i \pi_{A}^{(0)i} \bigg],  \label{H.c:final}
\end{align}
where $\tilde\sigma = \sum_{n=0}^\infty \sigma^{(n)}$. Then, the
primary Hamiltonian can be written as
\begin{align}
  \tilde{H}_{\mathrm{p}} = \tilde{H}_{\mathrm{c}}+ \int \dd[3]{\mathbf{x}} \left[\sum_{n=0}^\infty (u_1^{(n)}
  \Phi_1^{(n)} +u_2^{(n)} \Phi_2^{(n)}  +u_{3i}^{(n)}
  \Phi_{3}^{(n)i} ) + u_4^{(0)} \Phi_4^{(0)} \right], \label{H.p:final}
\end{align}
where $u$'s are the Lagrange multipliers related to the corresponding
primary constraints.  
Requiring the primary constraints to be preserved in time yields secondary constraints as
$\Phi_{5i} = -\dot{\Phi}_{3i}^{(n)} + \sum_{m=0}^\infty
\partial_i \Phi_2^{(m)} = \partial_i \pi_\chi^{(0)0} \approx 0$ and
$\Phi_6 = \dot{\Phi}_4 = \{\Phi_4, \tilde{H}_{\mathrm{p}}\} \approx 0$.
It is straightforward to verify that all primary and secondary constraints are
first-class.
After discarding total divergences, the action~\eqref{S:final:1st.class} can be written in a compact form as
\begin{align}
  \tilde{S} &= \int \dd[4]{x} \left[ - \frac14 \tilde\sigma \left(F^{(0)}_{\mu\nu}
      F^{(0)\mu\nu} - \partial_\mu \tilde\chi^{\mu} \right)\right], \label{S:final:simple}
\end{align}
where $\tilde \chi^\mu = \sum_{n=0}^\infty \chi^{(n)\mu}$.
The apparent form of the final action~\eqref{S:final:simple} resembles the starting action~\eqref{eq:action}, but the tilde fields
are defined as infinite sums of auxiliary fields.

\section{Local gauge symmetries}
\label{sec:new.symmetry}
The extended action~\eqref{S:final:simple} describes a first-class constraint system endowed with
local symmetries. To find the corresponding gauge transformation
rules, let us consider the transformation generator defined by
\begin{align}
  G = \int \dd[3]{\mathbf{x}} \left[ \sum_{n=0}^\infty \left( \epsilon_1^{(n)} \Phi_1^{(n)}+
  \epsilon_2^{(n)} \Phi_2^{(n)}+  \epsilon_{3i}^{(n)}
  \Phi_{3}^{(n)i}\right) + \epsilon_4 \Phi_4  + \epsilon_{5}^{i}
  \Phi_{5i} + \epsilon_6 \Phi_6 \right], \label{G:symmetry.generator}
\end{align}
where $\epsilon$'s are local gauge parameters.
The transformation of any phase-space variable is given by $\delta \phi = \{\phi, G\}$ for any field
$\phi$, and thus, the fields and momenta transform as
\begin{align}
  & \delta A_0^{(0)} = \epsilon_4, \quad \delta A^{(0)}_i = - \partial_i
    \epsilon_6, \quad \delta \pi_A^{(0)\mu} = 0, \label{delta.A:tmp}\\
  & \delta \sigma^{(n)} = \epsilon_1^{(n)} - \epsilon_1^{(n-1)}, \quad
    \delta \pi_\sigma^{(n)} = \frac14 \epsilon_2^{(n)}, \label{delta.sigma:tmp}\\
  & \delta \chi_0^{(n)} = \epsilon_2^{(n)} - \epsilon_2^{(n-1)} -
    \delta_{n0} \partial_i \epsilon_5^i, \quad \delta \pi_\chi^{(n)0}
    = - \frac14 \epsilon_1^{(n-1)}, \label{delta.chi.0:tmp}\\
  & \delta \chi_i^{(n)} = \epsilon_{3i}^{(n)}, \quad \delta
    \pi_\chi^{(n)i} = 0, \label{delta.chi.i:tmp}
\end{align}
where we set $\epsilon_a^{(-1)} = 0$.

Note that the secondary constraints $\Phi_{5i}$ and $\Phi_6$ arise from the time
evolution of the primary constraints. It means that the gauge parameters
corresponding to the secondary constraints are not independent but are related to
those of the primary constraints. Thus, imposing the condition
$\delta \dot{A}_i^{(0)} = \partial_0 (\delta A_i^{(0)})$ for the gauge field
leads to the relation $\epsilon_4 = -\dot{\epsilon}_6$.  Introducing a gauge
parameter $\lambda$ defined by $\lambda = -\epsilon_6$, the gauge transformation
of the vector potential in Eq.~\eqref{delta.A:tmp} reduces to the familiar form of
$\delta A_\mu^{(0)} = \partial_\mu \lambda$. In addition, the combined
condition
$\partial_{0} \Bigl( \delta \pi_{\sigma}^{(0)} - \tfrac{1}{4} \delta
\chi_{0}^{(0)} \Bigr) =\delta \Bigl( \dot{\pi}_{\sigma}^{(0)} - \tfrac{1}{4}
\dot{\chi}_{0}^{(0)} \Bigr)$ yields the relation of
$\sum_{n=0}^{\infty} \partial_{i} \epsilon_{3}^{(n)i} = \partial_{i}
\dot{\epsilon}_{5}^{i}$.  To satisfy the relation, we choose
$\epsilon_{3}^{(n)i} = \partial^{i} (\alpha^{(n)} - \alpha^{(n-1)}) +
\delta_{n0} \partial_{\mu} \Lambda^{i\mu}$ and $\epsilon_{5}^{i} = -\Lambda^{0i}$ along with $\epsilon_{2}^{(n)} = \dot{\alpha}^{(n)}$,
where $\alpha^{(n)}$ is a gauge
parameter satisfying $\alpha^{(-1)} = 0$.
In Eq.~\eqref{delta.sigma:tmp}, $\epsilon_1^{(n)} $ is also replaced by $ \epsilon^{(n)}$ for simplicity.
Then, the
gauge transformations of all fields in Eqs.~\eqref{delta.A:tmp}-\eqref{delta.chi.i:tmp} can be written as
\begin{gather}
  \delta A_\mu^{(0)} = \partial_\mu \lambda, \quad
  \delta \sigma^{(n)} = \epsilon^{(n)} - \epsilon^{(n-1)}, \quad
  \delta\chi^{(n)\mu} =  \delta_{n0}\, \partial_\nu \Lambda^{\mu\nu} +
  \partial^\mu (\alpha^{(n)} -
  \alpha^{(n-1)}), \label{gauge.transf:fields}
 \end{gather}
where $\lambda$, $\epsilon^{(n)}$, $\Lambda^{\mu\nu}$, and $\alpha^{(n)}$ are independent gauge parameters.
By imposing unitary gauges, $\sigma^{(n)}=0$ and $\chi^{(n)\mu}=0$
($n=1,2,\cdots$), through an appropriate choice of the gauge parameters of $\epsilon^{(n)}$ and $\alpha^{(n)}$, we recover the original action~\eqref{eq:action}, in which all auxiliary fields introduced in the course of the \ac{BFT} embedding vanish.
Consequently, the original action~\eqref{eq:action} can be regarded as the gauge-fixed version of
the extended action~\eqref{S:final:simple}.

\section{Conclusion and Discussion}
\label{sec:conclusion}

The Maxwell coupling constant was promoted to a scalar field $\sigma$,
accompanied by an auxiliary vector field $\chi^\mu$. The equation of motion for
$\sigma$ admits the constant solution,
which corresponds to a Noether charge associated with the gauge symmetry of
$\chi^\mu$. However, the Hamiltonian analysis reveals that certain gauge symmetries are broken, in the sense that a subset of the constraints becomes second-class.
Using the \ac{BFT}
formalism, we restored the gauge symmetries by converting the second-class
constraints into first-class ones in the extended phase space.  When both the
scalar field $\sigma$ and the auxiliary vector field $\chi^\mu$ are promoted to
infinite towers of fields, the resulting theory becomes
completely gauge invariant at the level of the action.

In the unitary gauge $\sigma^{(n)}=0$ and $\chi^{(n)\mu}=0$  for $n=1,2,\cdots$,
one can compute the Dirac brackets by reducing the degrees of freedom through additional gauge-fixing conditions.
For this purpose, one can choose the temporal and Coulomb gauges:
$\chi^0 \approx 0$ and $\partial_i A^i \approx 0$.
Requiring the time evolution of these constraints
yields the secondary constraints:
$-\frac{2}{\sigma^2} \pi_{A}^{(0)i} \pi_{A}^{(0)i} + F^{(0)}_{ij} F^{(0)ij} -
             \partial_i \chi^{(0)i} \approx 0$
and
$ -\partial_i \left( \frac{1}{\sigma} \pi_{A}^{(0)i} \right) +
                \partial_i \partial^i A_0^{(0)} \approx 0$.
The subsequent time evolution of the secondary constraints
determines all the remaining undetermined multiplier fields.
Therefore, the Dirac brackets can be explicitly constructed in the fully gauge-fixed system.


One might wonder whether the extended system~\eqref{S:final:simple}
admits any conserved charges arising from the extended symmetries or not.
From a technical standpoint, the extended local symmetries do not generate any new conserved charges, since they possess no nontrivial global components, as is evident from the choice of unitary gauge, which eliminates all auxiliary fields.

A final comment is in order.
Although we restricted our attention to the flat limit, it is natural to expect that a similar constraint structure in associated with the gauge symmetries may appear in the Einstein–Maxwell theory with the dynamical gauge coupling.
We hope that the present work provides a useful starting point for such studies.

\acknowledgments
This research was supported by Basic Science Research Program
through the National Research Foundation of Korea (NRF) funded by the Ministry
of Education through the Center for Quantum Spacetime (CQUeST) of Sogang
University (No. RS-2020-NR049598).  This work was supported by the National
Research Foundation of Korea (NRF) grant funded by the Korea government (MSIT)
(No. RS-2022-NR069013).


\bibliographystyle{JHEP}       

\bibliography{reference}

\end{document}